\documentclass[reprint, superscriptaddress, secnumarabic, amssymb, nobibnotes, aps, prl]{revtex4-1}

\usepackage{amsmath}
\usepackage{graphicx}
\usepackage{epstopdf}
\usepackage[]{natbib}
\usepackage{siunitx}

\begin{document}

\title{\textrm{Superconducting Properties and $\mu$SR Study of the Noncentrosymmetric Superconductor Nb$_{0.5}$Os$_{0.5}$}}
\author{D. Singh}
\affiliation{Department of Physics, Indian Institute of Science Education and Research Bhopal, Bhopal, 462066, India}
\author{J. A. T. Barker}
\affiliation{Physics Department, University of Warwick, Coventry CV4 7AL, United Kingdom}
\author{A. Thamizhavel}
\affiliation{Department of Condensed Matter Physics and Materials Science, Tata Institute of Fundamental Research, Mumbai 400005, India}
\author{A. D. Hillier}
\affiliation{ISIS Facility, STFC Rutherford Appleton Laboratory, Harwell Science and Innovation Campus, Oxfordshire, OX11 0QX, UK}
\author{D. McK. Paul}
\affiliation{Physics Department, University of Warwick, Coventry CV4 7AL, United Kingdom}
\author{R. P. Singh}
\email[]{rpsingh@iiserb.ac.in}
\affiliation{Department of Physics, Indian Institute of Science Education and Research Bhopal, Bhopal, 462066, India}

\date{\today}
\begin{abstract}
\begin{flushleft}
\end{flushleft}
The properties of the noncentrosymmetric superconductor ($\alpha$-$\textit{Mn}$ structure)  Nb$_{0.5}$Os$_{0.5}$ is investigated using resistivity, magnetization, specific heat, and muon spin relaxation and rotation ($\mu$SR) measurements. These measurements suggest that Nb$_{0.5}$Os$_{0.5}$ is a weakly coupled ($\lambda_{e-ph}$ $\sim$ 0.53) type-II superconductor ($\kappa_{GL}$ $\approx$ 61) having a bulk superconducting transition temperature $T_c$ = 3.07 K. The specific heat data in the superconductive regime fits well with the single-gap BCS model indicating nodeless s-wave superconductivity in Nb$_{0.5}$Os$_{0.5}$. The $\mu$SR measurements also confirm $\textit{s}$-wave superconductivity with the preserved time-reversal symmetry.
\end{abstract}
\maketitle
\section{Introduction}
Understanding the mechanism of unconventional superconductivity, where the structure lacks an inversion symmetry has been a tough challenge ever since the discovery of the heavy fermion noncentrosymmetric (NCS) superconductor CePt$_{3}$Si \cite{Bauer2004,EBA}. The lack of an inversion center in the crystal structure of the noncentrosymmetric superconductor makes parity an unconserved quantity. As a result, the superconducting ground state of an NCS superconductor may exhibit a possible mixing of spin-singlet and spin-triplet pair states \cite{rashba,sky,kv,ia,pa,fujimoto1,fujimoto2,fujimoto3,mdf}. The parity mixed superconducting ground state gives rise to several anomalous superconducting properties, e.g. upper critical field exceeding the Pauli limit, nodes in the superconducting gap, a helical vortex state, and time-reversal symmetry breaking.\\
Several NCS superconducting systems have been investigated to study the effects of broken inversion symmetry \cite{nr1,RT1,rf,rhf,rz3,YC,LC,rw,mib,lip,mac1,rg,ig,lrs,lps,rb1}, but majority of them appear to show $\textit{s}$-wave superconductivity. Theoretical predictions suggest that NCS superconductors are prime candidates to exhibit time-reversal symmetry breaking (TRSB) due to its admixed superconducting ground states. To date only a few NCS superconductors Re$_{6}$Zr \cite{rz1}, LaNiC$_{2}$ \cite{lnc2}, SrPtAs \cite{SPA} and La$_{7}$Ir$_{3}$ \cite{li1} have been reported to show TRSB. It is a rarely observed phenomena and apart from NCS superconductors, it has only been observed in a few unconventional superconductors e.g. Sr$_{2}$RuO$_{4}$ \cite{sro1,sro2}, UPt$_{3}$ \cite{UP1,UP2}, PrPt$_{4}$Ge$_{12}$ \cite{ppg}, LaNiGa$_{2}$ \cite{LNG}, and Lu$_{5}$Rh$_{6}$Sn$_{18}$ \cite{LoS}. The discrepancy between theory, experiment and the possibility of realizing an unconventional superconducting state having TRSB in NCS superconductors are of great interest. To understand the superconducting mechanism, it is required to study new NCS superconducting systems by combining bulk measurements such as transport, magnetization, heat capacity, etc. and local probe techniques like muon spectroscopy. Muon spectroscopy is one of the most direct methods of detecting the unconventional superconducting ground state. This technique can accurately determine the temperature dependence of the magnetic penetration depth and the onset of time-reversal symmetry breaking in superconductors.\\
Here we are reporting the superconducting state of a binary NCS compound ($\alpha$ - $Mn$ structure) Nb$_{0.5}$Os$_{0.5}$, having  superconducting transition $T_{c}$ = 3.07 K. Resistivity, magnetization, and specific heat measurements were carried out to explore the superconducting properties of Nb$_{0.5}$Os$_{0.5}$. $\mu$SR measurements in transverse-field (TF) and longitudinal-field (LF) configurations are used to probe the flux line lattice (FLL) and time-reversal symmetry breaking respectively.
\section{Experimental Details}
The polycrystalline sample of Nb$_{0.5}$Os$_{0.5}$ was prepared by arc melting. The stoichiometric amounts of Nb (99.95$\%$, Alfa Aesar) and Os (99.95$\%$, Alfa Aesar) were placed on the water cooled copper hearth in an ultrapure argon gas atmosphere. The sample was inverted and remelted several times to ensure sample homogeneity and the observed weight loss is negligible. The phase analysis was done using x-ray diffraction (XRD) at room temperature on a X'pert PANalytical diffractometer. The magnetization and ac susceptibility measurements were performed using the magnetic property measurement system (MPMS 3, Quantum Design Inc.). The electrical resistivity and specific heat measurements were done using the physical property measurement system (PPMS, Quantum Design Inc.).
The $\mu$SR measurements were carried out using the MuSR spectrometer at the ISIS facility, Rutherford Appleton Laboratory, Didcot, U. K. in both longitudinal and transverse geometries.
\begin{figure}
\includegraphics[width=1.0\columnwidth]{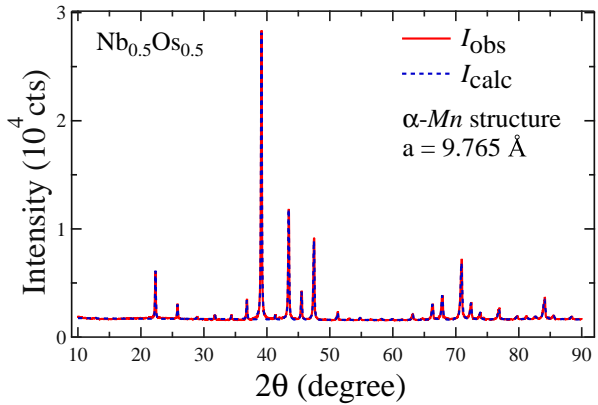}
\caption{\label{Fig1:xrd} Powder XRD pattern for the Nb$_{0.5}$Os$_{0.5}$ sample recorded at room temperature using Cu $K_{\alpha}$ radiation. The solid red line shows the experimental data. The dotted blue line corresponds to Rietveld refinement to the pattern.}
\end{figure}
\section{Results and Discussion}
\subsection{Sample characterization}
The powder x-ray diffraction pattern for Nb$_{0.5}$Os$_{0.5}$ was collected at room temperature. Rietveld refinement was performed using the High Score Plus Software. As observed from Fig. 1, the Nb$_{0.5}$Os$_{0.5}$ sample has no impurity phase. It can be indexed by cubic, noncentrosymmetric $\alpha$ - $Mn$ structure (space group $I \bar{4}3m$, No. 217) with the lattice cell parameter a = 9.765(3) \text{\AA}.
\subsection{Normal and superconducting state properties}
\subsubsection{Electrical resistivity}
\begin{figure}
\includegraphics[width=1.0\columnwidth]{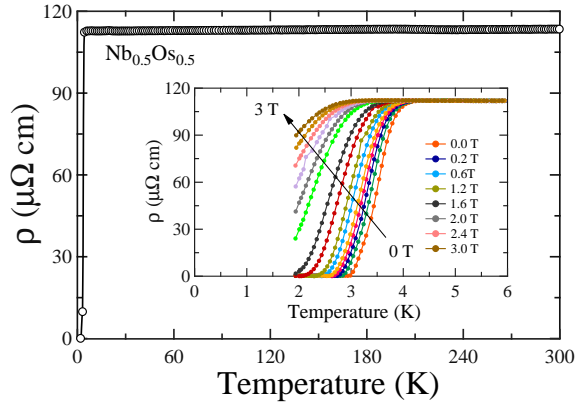}
\caption{\label{Fig2:Resistivity} The resistivity measurement $\rho(T)$ for Nb$_{0.5}$Os$_{0.5}$ taken in zero field in a temperature range of 1.85 K $\le$ T $\le$ 300 K. The inset shows $\rho(T)$ measurements as a function of magnetic fields.}
\end{figure}
The electrical resistivity measurement was done by the ac transport technique in the temperature range of 1.85 K $\le$ $\textit{T}$ $\le$ 300 K in zero field (see Fig. 2). The zero resistivity is acquired around $T_{c}^{0}$ $\approx$ 3.1 K. The normal state resistivity remains almost temperature independent up to the highest measured temperature, indicating that Nb$_{0.5}$Os$_{0.5}$ exhibit poor metallicity. The low value of the residual resistivity ratio (RRR) ($\frac{\rho(300)}{\rho(10)}$ = 1.05) suggests the dominance of strong electronic scattering due to the disorder. The resistivity measurements as a function of temperature were also done under different applied magnetic fields (up to 3T, see inset of Fig. 2) to calculate the higher critical field.
\subsubsection{Magnetization}
The magnetization measurement was done in zero-field cooled warming (ZFCW) and field cooled cooling (FCC) mode in an applied field of 5 mT (see Fig. 3(a)). The superconducting transition temperature was observed around $T_{c}^{onset}$ = 3.07 K, with the transition width of $\Delta T_{c}$ = 0.21 K. Low field M-H measurements were done at different temperatures to determine the lower critical field $H_{c1}$(0). It is defined as the first deviation from linearity in low-field regions in M vs H curves (see Fig. 3(b)). Using the formula $H_{c1}(T)= H_{c1}(0)(1-(T/T_{c})^{2})$ for the temperature variation of $H_{c1}(T)$, we estimated $H_{c1}$(0) = 3.06 $\pm$ 0.05 mT.
\begin{figure*}[t]
\includegraphics[width=2.1\columnwidth]{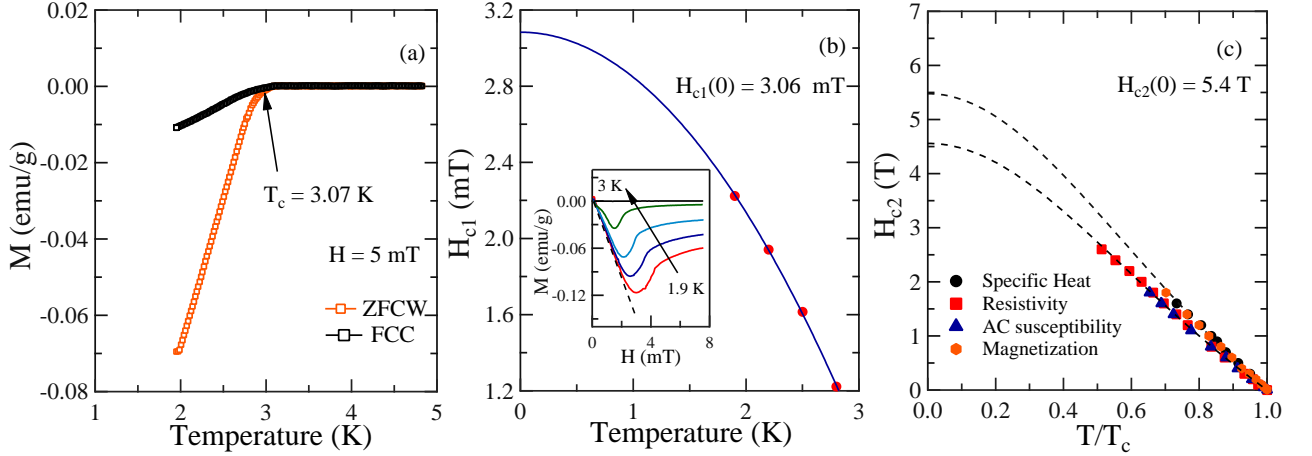}
\caption{\label{Fig3:zfc} (a) The magnetization data for Nb$_{0.5}$Os$_{0.5}$ taken in 5 mT field shows the superconducting transition at $T_{c}$ = 3.07 K. (b) The lower critical field $H_{c1}$ estimated by the GL formula was 3.06 mT. Inset shows the M vs H curves taken at various temperatures. (c) The upper critical field $H_{c2}$(T) obtained from magnetization, ac susceptibility, resistivity, and specific heat measurements. The dotted lines show the GL fits, yielding $H_{c2}$(0) $\simeq$ 5.4 T for Nb$_{0.5}$Os$_{0.5}$.} 
\end{figure*}

The temperature dependence of the upper critical field $H_{c2}$(T) was obtained by measuring the field dependence of superconducting transition $T_{c}$ in magnetization, ac susceptibility, resistivity, and specific heat measurements. It is evident from the graph (see Fig. 3(c)) that $H_{c2}$ varies linearly with the temperature and possibly best be fitted by the Ginzburg-Landau (GL) relation $H_{c2}(T) = H_{c2}(0)\frac{(1-t^{2})}{(1+t^2)}$, where t = $T/T_{c}$. By fitting above equation in the $H_{c2}$-T graph, the specific heat and magnetization measurements give $H_{c2}$(0) $\simeq$ 5.4 $\pm$ 0.1 T, whereas resistivity and ac susceptibility measurements give $H_{c2}$(0) $\simeq$ 4.6 $\pm$ 0.1 T. Using the relation $H_{c2}$(0) = $\Phi_{0}/2\pi\xi_{GL}^{2}$ where $\Phi_{0}$ is the quantum flux $(h/2e)$, we obtained $\xi_{GL}(0)$ = 78.12 \text{\AA}. Other superconducting parameters such as the Ginzburg Landau parameter $\kappa_{GL}$(0) (= 61), penetration depth $\lambda_{GL}$(0) (= 4774 \text{\AA}) and the thermodynamic critical field $H_{c}$(0) (= 62.6 mT) were calculated using the standard relations given in Ref. \cite{tin}.\\
For a type-II BCS superconductor in the dirty limit, the orbital limit of the upper critical field $H_{c2}^{orbital}$(0) is given by the Werthamer-Helfand-Hohenberg (WHH) \cite{EH,NRW} expression $H_{c2}^{orbital}$(0) = -0.693 $T_{c}\left.\frac{-dH_{c2}(T)}{dT}\right|_{T=T_{c}}$.
Using initial slope 2.1 T K$^{-1}$ from the $H_{c2}$-T phase diagram, $H_{c2}^{orbital}$(0) in the dirty limit was estimated to be 4.46 T. Within the $\alpha$-model the Pauli limiting field is given by $H_{c2}^{p}$(0) = 1.86$T_{c}(\alpha/\alpha_{BCS})$ \cite{DC}. Using $\alpha$ = 1.81 (from the specific heat measurement), it yields $H_{c2}^{p}$(0) = 5.85 T. The upper critical field $H_{c2}$(0) calculated above is close to both the orbital limiting field and Pauli limiting field. Therefore, it is highly desirable to perform the detailed investigations of the upper critical field in high quality single crystals of Nb$_{0.5}$Os$_{0.5}$.\\ 

 \subsubsection{Specific heat}
The temperature dependence of the specific heat was collected in zero field. The normal state low temperature specific heat data above $T_{c}$ can be fitted with the equation $C/T$ = $\gamma_{n}+\beta_{3}T^{2}+\beta_{5}T^{4}$ to the limit $\textit{T}$ $\to$ 0, to extract the electronic contribution ($\gamma_{n}$) and phononic contribution ($\beta_{3}$, $\beta_{5}$) to the specific heat. The solid red line in the inset of Fig. 4 shows the best fit to the data which yields $\gamma_{n}$ = 3.42 $\pm$ 0.01 mJ mol$^{-1}$ K$^{-2}$, $\beta_{3}$ = 0.039 $\pm$ 0.002 mJ mol$^{-1}$ K$^{-4}$, and $\beta_{5}$ = 0.205 $\pm$ 0.004 $\mu$J mol$^{-1}$ K$^{-6}$. The value of $\beta_{3}$ corresponds to a Debye temperature $\theta_{D}$ is 367 K. The Sommerfeld coefficient is proportional to the density of states $D_{C}(E_{F})$ at the Fermi level given by $\gamma_{n}$ = $(\pi^{2}k_{B}^{2}/3)D_{C}(E_{F})$, where using $\gamma_{n}$ = 3.42 $\pm$ 0.01 mJ mol$^{-1}$ K$^{-2}$ we obtained $D_{C}(E_{F})$ = 1.45 $\frac{states}{eV f.u}$.\\

\begin{figure}
\includegraphics[width=1.0\columnwidth]{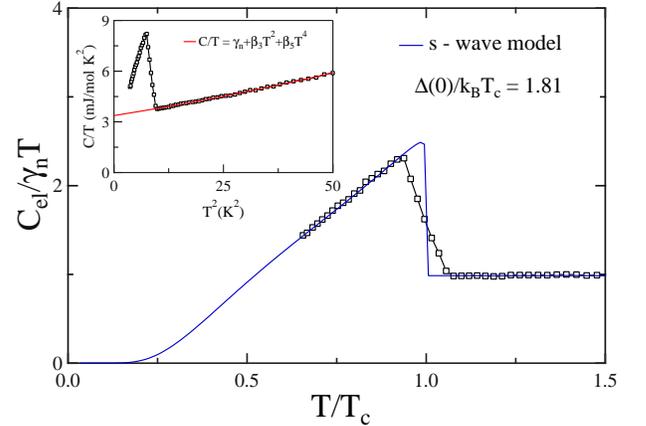}
\caption{\label{Fig4:hc2} Single-gap BCS expression given in Eq. (3) fits fairly well for $\Delta$(0)/$k_{B}T_{c}$ = 1.81 in Nb$_{0.5}$Os$_{0.5}$. Inset: The low temperature specific heat data above $T_{c}$ is fitted to the Debye model shown by solid red line.}
\end{figure}
 
The electron-phonon coupling constant can be calculated using the McMillan equation \cite{WL}
\begin{equation}
\lambda_{e-ph} = \frac{1.04+\mu^{*}ln(\theta_{D}/1.45T_{c})}{(1-0.62\mu^{*})ln(\theta_{D}/1.45T_{c})-1.04 } ,
\label{eqn1:ld}
\end{equation}                       
where $\mu^{*}$ is the Coulomb repulsion parameter, typically given by $\mu^{*}$ = 0.13 for many intermetallic superconductors. Using $T_{c}$=3.07 K and $\theta_{D}$ = 367 K for Nb$_{0.5}$Os$_{0.5}$, we obtained $\lambda_{e-ph}$ $\simeq$ 0.53. This value is comparable to other fully gapped NCS superconductors \cite{nr1,RT1,sas}, suggesting that Nb$_{0.5}$Os$_{0.5}$ is a weakly coupled superconductor. Using the value of $\lambda_{e-ph}$, we have calculated the effective mass for the quasiparticles $m^{*}$ = 1.53 $m_{e}$ \cite{GG}. The electronic contribution to the specific heat can be calculated by subtracting the phononic contribution. The normalized specific heat jump $\frac{\Delta C_{el}}{\gamma_{n}T_{c}}$ is 1.48 for $\gamma_{n}$ = 3.42 mJ mol$^{-1}$ K$^{-2}$, which is close to the value for a BCS superconductor (= 1.43) in the weak coupling limit.
The temperature dependence of the normalized entropy S in the superconducting state for a single-gap BCS superconductor is given by  
\begin{equation}
\frac{S}{\gamma_{n}T_{c}} = -\frac{6}{\pi^2}\left(\frac{\Delta(0)}{k_{B}T_{c}}\right)\int_{0}^{\infty}[ \textit{f}\ln(f)+(1-f)\ln(1-f)]dy ,
\label{eqn2:s}
\end{equation}
where $\textit{f}$($\xi$) = [exp($\textit{E}$($\xi$)/$k_{B}T$)+1]$^{-1}$ is the Fermi function, $\textit{E}$($\xi$) = $\sqrt{\xi^{2}+\Delta^{2}(t)}$, where $\xi$ is the energy of normal electrons measured relative to the Fermi energy, $\textit{y}$ = $\xi/\Delta(0)$, $\mathit{t = T/T_{c}}$, and $\Delta(t)$ = tanh[1.82(1.018(($\mathit{1/t}$)-1))$^{0.51}$] is the BCS approximation for the temperature dependence of the energy gap. The normalized electronic specific heat is then calculated from the normalized entropy by
\begin{equation}
\frac{C_{el}}{\gamma_{n}T_{c}} = t\frac{d(S/\gamma_{n}T_{c})}{dt} .
\label{eqn3:Cel}
\end{equation}
The $C_{el}$ below $T_{c}$ is described by Eq. (3) whereas above $T_{c}$ its equal to $\gamma_{n}T_{c}$. Figure 4 shows the fitting of the specific heat data using Eq. (3), which yields $\alpha$ = $\Delta(0)/k_{B}T_{c}$ = 1.81 $\pm$ 0.02. The obtained value is close to the BCS value $\alpha_{BCS}$ = 1.764 in the weak coupling limit, suggesting single-gap BCS like superconductivity in Nb$_{0.5}$Os$_{0.5}$.\\  
In the $\alpha$ model, BCS parameter $\alpha_{BCS}$ is replaced by $\alpha$ which can be determined using the formula $\Delta C_{el}/\gamma_{n}T_{c} = 1.426(\alpha/\alpha_{BCS})^{2}$ \cite{DC}. Substituting the value of normalized specific heat jump $\Delta C_{el}/\gamma_{n}T_{c}$ = 1.48 for our sample, we get $\alpha$ = 1.8, which is in good agreement with the fitted value.
\\

\begin{table}[h!]
\caption{Normal and superconducting properties of Nb$_{0.5}$Os$_{0.5}$}
\begin{center}
\begin{tabular}[b]{lcc}\hline\hline
Parameter& unit& value\\
\hline
\\[0.5ex]                                  
$T_{c}$& K& 3.07\\             
$H_{c1}(0)$& mT& 3.06 \\                       
$H_{c2}(0)$& T& 5.4 \\
$H_{c}(0)$& mT& 62.6 \\
$H_{c2}^{orbital}(0)$& T& 4.46\\
$H_{c2}^{P}(0)$& T& 5.85\\
$\xi_{GL}$& \text{\AA}& 78.12\\
$\lambda_{GL}$& \text{\AA}& 4774\\
$\kappa_{GL}$& &61\\
$\gamma$& mJmol$^{-1}$K$^{-2}$& 3.42\\
$\beta$ & mJmol$^{-1}$K$^{-4}$& 0.039\\
$\theta_{D}$& K& 367\\
$\lambda_{e-ph}$&  &0.53\\
D$_{C}$(E$_{f}$)& states/ev f.u& 1.45\\
$\Delta C_{el}/\gamma_{n}T_{c}$&   &1.48\\
$\Delta(0)/k_{B}T_{c}$&   &1.81
\\[0.5ex]
\hline\hline
\end{tabular}
\par\medskip\footnotesize
\end{center}
\end{table}
\subsubsection{Muon spin relaxation and rotation}

The superconducting ground state of Nb$_{0.5}$Os$_{0.5}$ was further analyzed by $\mu$SR relaxation and rotation measurements. The zero-field muon spin relaxation (ZF-$\mu$SR) spectra was collected below ($T$ = 40 mK) and above ($T$ = 3.5 K) the transition temperature ($T_{c}$ = 3.07 K) as displayed in Fig. 5. The absence of any oscillatory component in the spectra confirms that there are no atomic moments, generally associated with the ordered magnetic structure. In the absence of atomic moments, muon-spin relaxation in zero field is given by the Gaussian Kubo-Toyabe (KT) function \cite{RSH} 
\begin{equation}
G_{\mathrm{KT}}(t) = \frac{1}{3}+\frac{2}{3}(1-\sigma^{2}_{\mathrm{ZF}}t^{2})\mathrm{exp}\left(\frac{-\sigma^{2}_{\mathrm{ZF}}t^{2}}{2}\right) ,
\label{eqn4:zf}
\end{equation} 
where $\sigma_{\mathrm{ZF}}$ accounts for the relaxation due to static, randomly oriented local fields associated with the nuclear moments at the muon site.
The spectra well described by the function
\begin{equation}
A(t) = A_{1}G_{\mathrm{KT}}(t)\mathrm{exp}(-\Lambda t)+A_{\mathrm{BG}} ,
\label{eqn5:tay}
\end{equation} 
where $A_{1}$ is the initial asymmetry, $\Lambda$ is the electronic relaxation rate, and $A_{\mathrm{BG}}$ is the time-independent background contribution from the muons stopped in the sample holder. By fitting both the ZF-$\mu$SR spectra (Fig. 5 ) with the Eq. (5), yields the similar set of parameters within the sensitivity of the instrument. In the superconducting state, if the spin-triplet component is present, an additional relaxation should be observed \cite{lnc2,rz1,SPA,li1,sro1}. It is clearly absent in Fig. 5, where identical relaxation signals can be observed on the either side of the superconducting transition temperature. This leads to the conclusion that the time-reversal symmetry is preserved in Nb$_{0.5}$Os$_{0.5}$ within the detection limit of $\mu$SR.\\ 
\begin{figure}
\includegraphics[width=1.0\columnwidth]{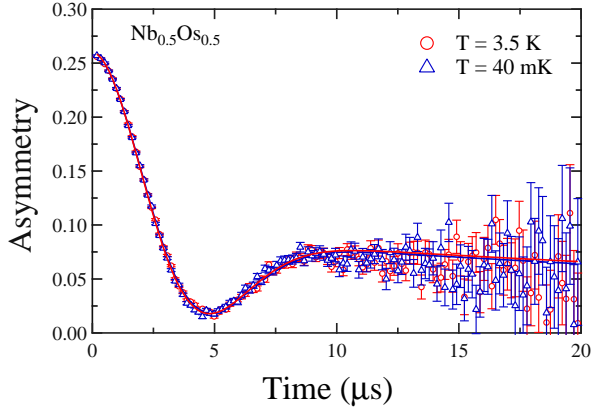}
\caption{\label{Fig5:ZFM} Zero field $\mu$SR spectra collected below (40 mK) and above (3.5 K) the superconducting transition temperature. The solid lines are the fits to Gaussian Kubo-Toyabe (KT) function given in Eq. (5). }
\end{figure}
\begin{figure}
\includegraphics[width=1.0\columnwidth]{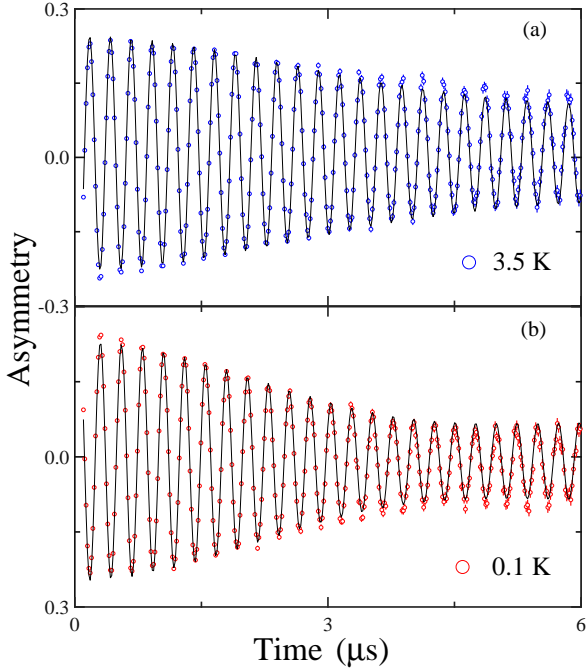}
\caption{\label{Fig6:TFM} Representative TF $\mu$SR signals collected at (a) 3.5 K and (b) 0.1 K in an applied magnetic field of 30 mT.
The solid lines are fits using Eq. (6).}
\end{figure} 
Transverse-field muon spin rotation (TF-$\mu$SR) measurements were done to gain information on the superconducting gap structure of Nb$_{0.5}$Os$_{0.5}$. Asymmetry spectra was recorded above (3.5 K) and below (0.1 K) the transition temperature $T_{c}$ in a transverse field of 30 mT as shown in Fig. 6. The TF-$\mu$SR precession signal were fitted using an oscillatory decaying Gaussian function
\begin{equation}
G_{\mathrm{TF}}(t) = A_{1}\mathrm{exp}\left(\frac{-\sigma^{2}t^{2}}{2}\right)\mathrm{cos}(w_{1}t+\phi)+A_{2}\mathrm{cos}(w_{2}t+\phi) ,
\label{eqn6:Tranf}
\end{equation}
where $w_{1}$ and $w_{2}$ are the frequencies of the muon precession signal and background signal respectively, $\phi$ is the initial phase offset and $\sigma$ is the Gaussian muon-spin relaxation rate. Figure 6(a) shows the signal in the normal state where depolarization rate is small, attributed to homogeneous field distribution throughout the sample. The significant depolarization rate in the superconducting state shown in the Fig. 6(b) is due to the flux line lattice (FLL) in the mixed state of the superconductor, which gives rise to the inhomogeneous field distribution. The depolarization arising due to the static fields from the nuclear moments $\sigma_{\mathrm{N}}$ is assumed to be temperature independent and adds in quadrature to the contribution from the field variation across the flux line lattice $\sigma_{\mathrm{FLL}}$:
\begin{equation}
\sigma^{2} = \sigma_{\mathrm{N}}^{2}+\sigma_{\mathrm{FLL}}^{2} .
\label{eqn7:sigma}
\end{equation}

The muon-spin relaxation rate in the superconducting state $\sigma_{\mathrm{FLL}}$ is related to the London magnetic penetration depth $\lambda$ and thus to the superfluid density $n_{s}$ by the equation 

\begin{equation}
\frac{\sigma_{\mathrm{FLL}}(T)}{\sigma_{\mathrm{FLL}}(0)} = \frac{\lambda^{-2}(T)}{\lambda^{-2}(0)} .
\label{eqn8:sfd}
\end{equation}

For an $\textit{s}$-wave BCS superconductor in the dirty limit, the temperature dependence of the London magnetic penetration depth is given by 
\begin{equation}
\frac{\lambda^{-2}(T)}{\lambda^{-2}(0)} = \frac{\Delta(T)}{\Delta(0)}\mathrm{tanh}\left[\frac{\Delta(T)}{2k_{B}T}\right] ,
\label{eqn9:lpd}
\end{equation}
 where $\Delta$(T) = $\Delta_{0}$$\delta(T/T_{c})$. The temperature dependence of the gap in the BCS approximation is given by the expression $\delta(T/T_{c})$ = tanh[1.82(1.018($\mathit{(T_{c}/T})$-1))$^{0.51}$]. Taking the dirty limit expression for Nb$_{0.5}$Os$_{0.5}$ and  combining Eq. (7), (8) and (9), a model was obtained for a dirty limit single-gap s-wave superconductor, where $\sigma$(T) above ${T}_{c}$ is equal to $\sigma_{\mathrm{N}}$ and below ${T}_{c}$ is given by Eq. (10) which contain contributions from both $\sigma_{\mathrm{N}}$ and $\sigma_{\mathrm{FLL}}$.

\begin{equation}
\sigma(T) = \sqrt{\sigma_{\mathrm{FLL}}^{2}(0)\frac{\Delta^{2}(T)}{\Delta^{2}(0)}\mathrm{tanh}^{2}\left[\frac{\Delta(T)}{2k_{B}T}\right]+\sigma_{\mathrm{N}}^{2}}  .
\label{eqn10:fs}
\end{equation}
 
\begin{figure}
\includegraphics[width=1.0\columnwidth]{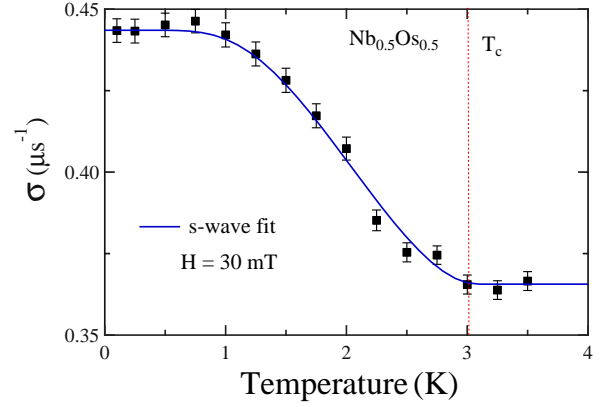}
\caption{\label{Fig7:swave} The temperature dependence of the muon-spin relaxation rate $\sigma$(T) collected at an applied field of 30 mT. The solid blue line shows the s-wave fit for a dirty limit superconductor using Eq. (10).}
\end{figure}

The temperature dependence of muon depolarization rate $\sigma$ was collected in an applied field of 30 mT as shown in Fig. 7. The depolarization rate $\sigma$ remains temperature independent up to $T_{c}$ attributing to random nuclear magnetic moments, then after $T_{c}$, $\sigma$ increases due to the formation of well-ordered FLL. The best fit to the $\sigma(T)$ data were obtained with the single-gap BCS model (Eq. (10)) shown by the solid blue line in Fig. 7, where we have obtained $\sigma_{\mathrm{N}}$ = 0.366 $\pm$ 0.002 $\mu$s$^{-1}$, $\sigma$(0) = 0.444 $\pm$ 0.001 $\mu$s$^{-1}$, and $\Delta$(0) = 0.50 $\pm$ 0.02 meV. The value of $\alpha$ = $\Delta(0)/k_{B}T_{c}$ = 1.89 is close to the value ($\alpha$ = 1.81) obtained from the low temperature specific heat measurement. Thus, the TF- $\mu$SR measurements together with the specific heat measurement confirm that Nb$_{0.5}$Os$_{0.5}$ is a s-wave superconductor.\\
The penetration depth $\lambda$(0) at T = 0 K can be directly calculated ($\sigma_{\mathrm{FLL}}$(0) = 0.251 $\pm$ 0.001 $\mu$s$^{-1}$) from the relation \cite{JES,EHB}
\begin{equation}
\frac{\sigma_{\mathrm{FLL}}^2(0)}{\gamma_{\mu}^2} = 0.00371 \frac{\Phi_{0}^{2}}{\lambda^{4}(0)} ,
\label{eqn11:lam}
\end{equation} 
where $\gamma_{\mu}$/2$\pi$ = 135.53 MHz/T is the muon gyromagnetic ratio and $\Phi_{0}$ is the magnetic flux quantum. The value of penetration depth $\lambda$(0) is 6538$\pm$13 \text{\AA}. The estimated value is little higher than the $\lambda_{GL}$(0), which could be due to the dirty limit superconductivity in Nb$_{0.5}$Os$_{0.5}$.\\
Uemura $\textit{et al.}$ showed in 1991 that the superconductors can be classified into a conventional/unconventional superconductor \cite{KK,YJU} based on the ratio of the transition temperature ($T_{c}$) to the Fermi temperature ($T_{F}$). It was shown that the unconventional, exotic superconductors fall in the range of 0.01 $\leq$ $\frac{T_{c}}{T_{F}}$ $\leq$ 0.1.
\begin{figure}
\includegraphics[width=1.0\columnwidth]{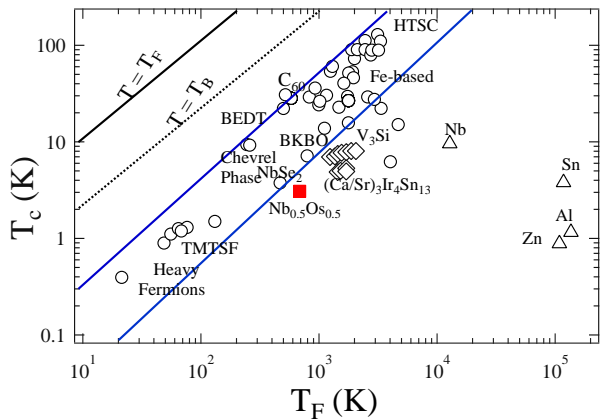}
\caption{\label{Fig8:up} The Uemura plot showing the superconducting transition temperature $T_{c}$ vs the effective Fermi temperature $T_{F}$, where Nb$_{0.5}$Os$_{0.5}$ is shown as a solid red square. Other data points plotted between the blue solid lines is the different families of unconventional superconductors.}
\end{figure}
The Fermi temperature can be calculated using the relation 
\begin{equation}
k_{B}T_{F} = \frac{\hbar^{2}}{2}(3\pi^{2})^{2/3}\frac{n_{s}^{2/3}}{m_{e}[1+\lambda_{e-ph}]} ,
\label{eqn12:ft}
\end{equation}
where n$_{s}$ is the density of paired electrons and $\lambda_{e-ph}$ is the electron-phonon coupling constant.
Using the Sommerfeld coefficient for Nb$_{0.5}$Os$_{0.5}$ \cite{ck}, we have calculated the number density of electrons n$_{e}$ = 2.94 $\times$ 10$^{30}$ $m^{-3}$. The estimated value of $\textit{l}$ (0.56 \text{\AA}) $\ll$ $\xi_{0}$ (14091 \text{\AA}), means that in Nb$_{0.5}$Os$_{0.5}$ the density of paired electrons will be given by n$_{s}$ $\simeq$ n$_{e}$ $\frac{\textit{l}}{\xi_{0}}$ = 1.17 $\times$ 10$^{26}$ $m^{-3}$. The above result is verified from the magnetic penetration depth $\lambda$ calculated from the muon analysis, where the density of paired electrons is given by n$_{s}$ = $\frac{m_{e}(1+ \lambda_{e-ph})}{\mu_{0}e^{2}\lambda^{2}}$ $\simeq$ 1.01 $\times$ 10$^{26}$ $m^{-3}$.\\ 
Using the value of n$_{s}$ in Eq. (12), it yields $T_{F}$ = 662 K, giving the ratio $\frac{T_{c}}{T_{F}}$ = 0.0046, just outside the range of unconventional superconductors as shown by a solid red square in Fig. 8, where blue solid lines represent the band of unconventional superconductors. A similar result is obtained if we express the superfluid density in term of the muon spin-relaxation rate $\sigma(0)$ $\propto$ $\lambda(0)^{-2}$ $\propto$ $\rho_{s}$(0) as in the original Uemura plot.
\section{Conclusion}

The transport, magnetization, and heat capacity measurements confirm type-II, $\textit{s}$-wave superconductivity in Nb$_{0.5}$Os$_{0.5}$ having transition temperature $T_{c}$ = 3.07 K. The upper and lower critical fields estimated to be $H_{c1}$$\simeq$ 3.06 mT and $H_{c2}$$\simeq$ 5.4 T respectively. The TF-$\mu$SR measurements further confirm $\textit{s}$-wave superconductivity. The ZF-$\mu$SR measurements show no evidence of long-range magnetic ordering and any additional relaxation channel in the superconducting state. It confirms that time-reversal symmetry is preserved in Nb$_{0.5}$Os$_{0.5}$. This result contradicts the possibility of time-reversal symmetry breaking in NCS superconductors due to the admixed pairing states (spin-singlet/spin-triplet). Several other NCS superconductors (weakly/strongly correlated) reported to show the similar result. It suggest some other mechanism may be involved, which control the TRSB in NCS superconductors. In order to understand the presence and absence of time-reversal symmetry breaking in NCS superconducting compounds, it is clearly important to search the new NCS superconductor.

\section{Acknowledgments}

R.~P.~S.\ acknowledges Science and Engineering Research Board, Government of India for the Ramanujan Fellowship through Grant No. SR/S2/RJN-83/2012 and Newton Bhabha funding. We thank ISIS, STFC, UK for the muon beamtime to conduct the $\mu$SR experiments.

\end{document}